\documentclass[twocolumn,showpacs,prb]{revtex4-1}%

\usepackage{color}
\usepackage{amsmath}
\usepackage{amsfonts}
\usepackage{amssymb}
\usepackage{graphicx}%
\usepackage{epsfig}%
\usepackage{srcltx}%

\begin{document}
\title{
  Au$_{40}$: A Large Tetrahedral Magic Cluster
}
\author{De-en Jiang$^{1}$}
\author{Michael Walter$^{2}$}
\affiliation{$^1$ Chemical Sciences Division, Oak Ridge National
  Laboratory, Oak Ridge, TN 37831-6201}
\affiliation{$^2$ Freiburg Materials Research Center,
  Stefan-Meier-Str. 21, 79104 Freiburg, Germany}

\begin{abstract}
40 is a magic number for tetrahedral symmetry predicted in both
nuclear physics and the electronic jellium model. We show that
Au$_{40}$ could be such a a magic cluster from density functional
theory-based basin hopping for global minimization. The putative
global minimum found for Au$_{40}$ has a twisted pyramid structure,
reminiscent of the famous tetrahedral Au$_{20}$,
and a sizable HOMO-LUMO gap of
0.69 eV, indicating its molecular nature.
Analysis of the electronic states reveals that the gap is related to
shell closings of the metallic electrons in a tetrahedrally distorted
effective potential.
\end{abstract}

\maketitle


Gold is a magic element that constantly brings us
surprises. The relativistic effect has been demonstrated to be a
dominating factor that distinguishes gold from the same-group elements
such as copper and silver \cite{Hakkinen02prl}.
Many intriguing structures and properties
are found for nanometer-sized gold clusters. One example is the
transition from a two-dimensional structure to a three-dimensional one
around a size of 10 atoms \cite{Hakkinen08csr,Ferrighi09jacs}.
Another is the unique catalytic activity of
nanometer-sized gold clusters \cite{Yoon05sc}.

Among all gold nanoclusters, Au$_{20}$ is probably the most famous one
\cite{Li03sc}.
It has a simple tetrahedral symmetry with every gold atom on the
surface. The tetrahedral symmetry is very stable and is
the ground state both in charged and neutral clusters
\cite{Gruene08sc}. The symmetry is still retained if one atom
is removed to form Au$_{19}$ \cite{Gruene08sc} and is
proposed to be present in chemically synthesized
Au$_{20}$(PPh$_3$)$_8$ clusters \cite{Zhang04jpcb}.

It is common for gold clusters that the frontier orbitals
around Fermi energy are
derived from Au($6s$) atomic orbitals and form delocalized states
distributed over the whole cluster \cite{Yoon07cpc}.
Very similar to the stability of noble gas atoms
in the periodic system of elements, a closed electronic shell
built up from these delocalized states improves the
energetics and gives
more chemical stability, i.e. reduced reactivity \cite{Heer93rmp}.
This effect also leads to
enhanced stability for specific sizes in protected gold
clusters \cite{Walter08pnas,Walter11cs}.
Spherical shell closings
are especially prominent as these are known to lead to large gaps between
highest occupied (HOMO) and lowest unoccupied molecular orbitals
(LUMO), the signature of chemical stability.

In the case of
deformations of the background potential, the Jahn-Teller effect can
lead to stabilizations where spherical shell closings are not
available \cite{Heer93rmp}. A special case here is
octupole deformations,
where in particular tetrahedral deformations produce large gaps \cite{Hamamoto91zpd}.

The appearance of tetrahedral deformations
is proposed in nuclear physics \cite{Dudek02prl}
where the delocalized particle picture in an effective background potential
had originated before it was adopted in cluster physics.
While the experimental observation of
the tetrahedral deformation in nuclei seems to be under debate
\cite{Bark10prl,Jentschel10prl},
the importance of tetrahedral symmetry in cluster physics
in the case of Au$_{20}$ is beyond
question \footnote{
  Also tetrahedral [Os$_{10}$C(CO)$_{24}$]$^{2-}$ \cite{Johnson85nat} and
  [Os$_{20}$(CO)$_{40}$]$^{2-}$ \cite{Amoroso91ac} exist.
  Their stability is not related to electronic shell closings as indicated
  by their magnetism \cite{Johnson85nat}.
}.

Magic tetrahedral metal clusters were proposed in the jellium model
\cite{Hamamoto91zpd}. In particular, in the
essentially parameter-free ultimate
jellium model,
where the smeared out nuclear density exactly follows the electronic
density, the 40 electron ground state shows a tetrahedral
deformation \cite{Reimann97prb}.
In spherical symmetry the electrons' angular momentum
is conserved and the relative energies of different angular
momentum shells (and also the
gaps between them) depend on the effective radial potential.
There is a large gap for a harmonic radial potential
at 40 electrons.
This gap decreases when the potential becomes more box like, however
\cite{Heer93rmp}.
Here a tetrahedral deformation can open the
gap again. To our knowledge, tetrahedral metal clusters larger than the rather
trivial case of four atoms have been observed only in the example of
Au$_{20}$. In this Letter, we show from density functional theory-based global minimization that the larger Au$_{40}$ has a twisted trigonal
pyramid structure of quasi-tetrahedral
symmetry, the first case of a metal cluster of
tetrahedral shape beyond Au$_{20}$.


Although gold clusters with 20 atoms or less have been
extensively studied, we know relatively little of the structures of
larger gold clusters. Au clusters with 30 to 60 atoms would be the key to
understanding the transition from the molecular behavior of a small
cluster to the metallic bulk.
The clusters Au$_{32}$ \cite{Ji05ac,Jalbout08jpca}
and Au$_{34}$ \cite{Lechtken07ac} have been proposed to
have a core-shell structure, instead of being hollow or planar. More recently, the
global minima of Au$_{28}$ to Au$_{35}$ were explored in comparison with experimental
photoelectron spectra
\cite{Shao10jacs}.
The authors found that the global minima are amorphous in
nature with an Au$_4$ tetrahedron core and a much bigger outer-shell
for Au$_N$ with $N>32$.
Using an empirical potential for global minimum search followed by
density functional theory (DFT)
calculations, Garz\'on et al. found that the most stable structure
of Au$_{38}$ is of $C_s$ symmetry with an Au$_5$ core
\cite{Garzon98prl,Garzon03epjd}. This structure was
found to be slightly lower in energy than the high-symmetry
truncated octahedron (by about 0.3 to 0.6 eV, depending on the choice
of DFT functional \cite{Garzon98prl,Garzon03epjd}). Both the $C_s$ and
octahedral structures are metallic
(that is, their HOMOs are not completely filled).
Using a strategy similar to Garz\'on et al.'s,
Tran and Johnston found a structure for Au$_{40}$ with a
distorted truncated octahedron \cite{Tran11prs}.

\begin{figure}[ht]
  \epsfig{file=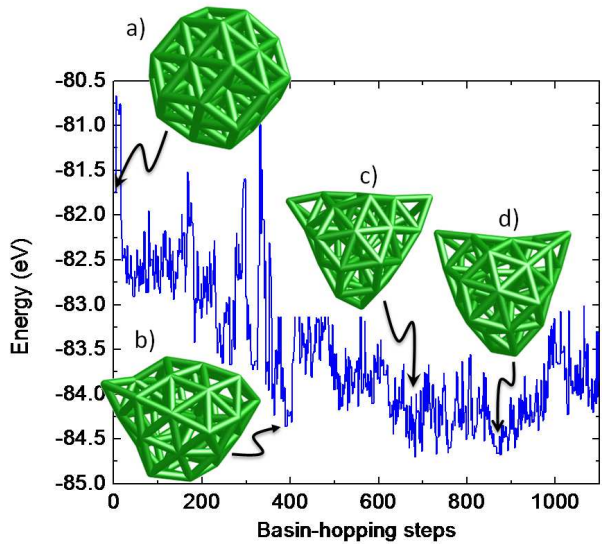, width=\linewidth}
  \caption{
    \label{fig:Fig1}
    Energy landscape of a DFT-based basin-hopping search for the global minimum of
    Au$_{38}$, started with the octahedral structure.
  }
\end{figure}

Puzzled by the metallic nature of
Au$_{38}$ of the state-of-the-art models
and expecting a high-symmetry structure for Au$_{40}$, we set out to find
the global minima of Au$_{38}$ and Au$_{40}$, hypothesizing that Au$_{40}$'s structure will build upon that of Au$_{38}$. Instead of using the empirical
potentials to search for global minima before DFT optimization, as
previously done for Au$_{38}$ and Au$_{40}$, we use DFT geometry optimization
at the GGA-PBE level
\footnote{Generalized gradient approximation (GGA) as devised by
Perdew, Burke and Ernzerhof \cite{Perdew96prl}.
Calculations using a higher level meta-GGA approximation \cite{Tao03prl}
yield similar results.}
directly in our basin-hopping global-minimum search
\footnote{A Python
  script was used to interface the basin-hopping algorithm
  \cite{Wales97jpca} and the Vienna Ab initio Simulation
  Package \cite{Kresse96prb,Kresse96cms} for planewave DFT calculations
  with the scalar-relativistic projector-augmented wave (PAW)
  potential for Au which allows use of a low cutoff energy of 172.5 eV. The neutral
  cluster was placed in the center of a $1.8\times1.8\times1.8$ nm$^3$
  box. At each step, a full geometry optimization was performed, and the
  final energy was compared with the previous one for a Metroplis
  sampling with a temperature of 7500 K. Then, all Cartesian
  coordinates were displaced by a random number in [-1,1] times the step size
  of 0.07 nm. After about 1000 steps which takes 256 parallel cores on
  a Cray XE6 supercomputer 67 hours to finish, the chosen candidate
  structures were then finely relaxed with a molecular quantumn
  chemistry code, Turbomole \cite{Ahlrichs89cpl} with the def2-TZVP
  basis sets.}.
This approach has
been quite powerful for exploring the energy landscape of
nanoclusters \cite{Jiang10cej,Jiang11acsnano}. What distinguishes our
work from previous DFT-based
basin-hopping search for the similar-sized gold clusters is that
we run the basin-hopping procedure for much more steps (over 1000) to explore the
energy landscape.

Fig. 1 shows our DFT-based basin-hopping search for the global minimum of Au$_{38}$, started with the highly symmetric octahedral structure. One can see that the octahedral
structure (Fig. 1a) was transformed into much less symmetric
configurations of lower energy. The energy lowering
is quite substantial, more than 2 eV,
instead of the
0.3 to 0.6 eV lowering found by Garz\'on et al. for their $C_s$ model
\cite{Garzon98prl,Garzon03epjd}.
After a local minimum with one adatom sticking out (Fig. 1b), we
found two putative global minima (Fig. 1c and 1d) which are almost
degenerate in energy (within 30 meV). The two structures have a
similar construction: an Au$_4$ core and an Au$_{32}$ shell, with two
Au adatoms sticking out; but one with $C_1$ symmetry
(Fig. 1c), the
other with $C_2$
symmetry (Fig. 1d).
What is unique about the two structures is that they both
have a sizable HOMO-LUMO gap, indicating their molecule-like stability
in the gas phase.
The $C_1$ structure has a gap of 0.66 eV and
the $C_2$ structure
0.84 eV, unlike the metallic nature of the octahedral structure and Garz\'on et al.'s
$C_s$ model \cite{Garzon98prl,Garzon03epjd}.
Hence we confirmed that the nanometer-sized Au$_{38}$
still behaves as a molecule.

\begin{figure}[ht]
  \epsfig{file=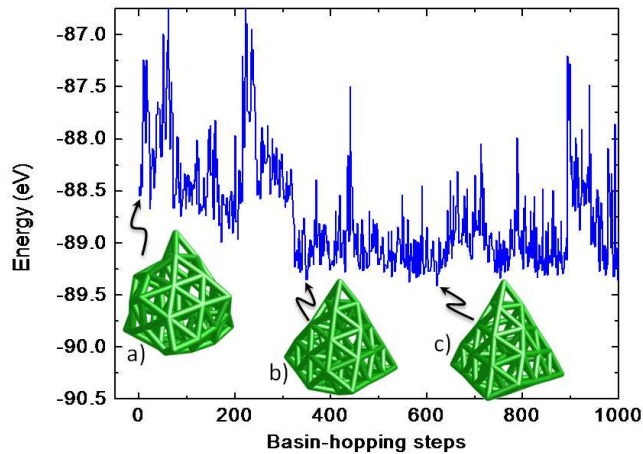, width=\linewidth}
  \caption{
    \label{fig:Fig3}
     Energy landscape of a DFT-based basin-hopping search for the global minimum
     of Au$_{40}$.
  }
\end{figure}

\begin{figure}[ht]
  \epsfig{file=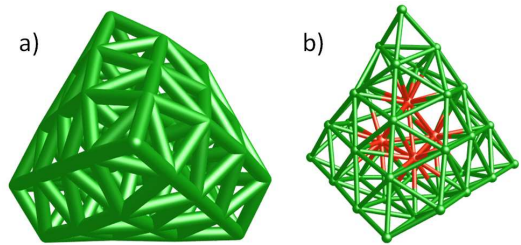, width=0.8\linewidth}
  \caption{
    \label{fig:Fig4}
     The global minimum for Au$_{40}$ featuring a twisted pyramid with a
     missing corner: (a) top view, the missing corner is at the top of
     the figure; (b) side view, the Au$_4$ tetrahedral core is
     highlighted in red.
  }
\end{figure}

The two Au$_{38}$ minima are also interesting in that
they have two adatoms on the Au$_4$@Au$_{32}$ core shell.
This construction shares some
similarity with the Au$_{35}$ structure where one
adatom is on the Au$_4$@Au$_{30}$
core shell \cite{Shao10jacs}. What Au$_{38}$'s structure suggested to
us is that Au$_{40}$'s
structure can build upon the Au$_4$@Au$_{32}$ core-shell framework but
with four adatoms placed in tetrahedral symmetry. This idea led us to propose
an initial guess for
Au$_{40}$ (Fig. 2a) with two more adatoms manually added to one of the
Au$_{38}$'s two candidate structures (Fig. 1c). Started with this
initial guess, we
performed DFT-based basin-hopping search for over 1000 steps. Interestingly, twisted pyramid structures evolved
out. The first one evolved out is a twisted trigonal pyramid with a
missing corner (Fig. 2b and Fig. 3); the core is still an Au$_4$
tetrahedron (Fig. 3b); overall, the cluster has $C_1$ symmetry. The second one
evolved out is a twisted trigonal pyramid with a missing core atom
(Fig. 2c and Fig. 4); this structure has $C_3$ symmetry, with an Au$_3$
triangle core (Fig. 4b). The $C_3$ symmetry can also be clearly seen from the base of the pyramid (Fig. 4c). The top three
layers of the pyramid share the same substructure as the famous
tetrahedral Au$_{20}$. Both Au$_{40}$ structures are chiral as
was first proposed
for Au$_{34}$ \cite{Lechtken07ac}.

\begin{figure}[ht]
  \epsfig{file=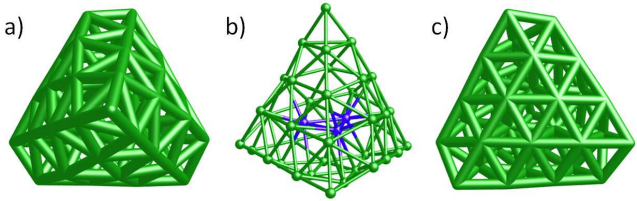, width=\linewidth}
  \caption{
    \label{fig:Fig5}
    An isomer of Au$_{40}$ featuring a twisted pyramid with $C_3$ symmetry: (a)
    top view; (b) side view, showing the Au$_3$ triangle core; (c) bottom
    view, showing the base of the pyramid.
  }
\end{figure}

To gain a deeper insight into the consequences
of the tetrahedral symmetry we
now discuss the electronic structure and energetics of the two Au$_{40}$
isomers. We found that the $C_1$ isomer is the most stable among all the
configurations explored;
it has a HOMO-LUMO gap of 0.69 eV. The $C_3$
isomer is only 0.15 eV higher in energy and has an even larger
gap of 0.85
eV. So both isomers are stable molecules in nature, given these
rather large gaps in clusters of this size.
Moreover, we found
that if one manually moves the apex atom of the $C_1$ isomer to the
missing corner (Fig. 3a), the resultant structure is only slightly
higher in energy (by 0.07 eV), indicating the robustness of
Au$_{40}$'s
tetrahedral shape despite the multiple isomers close in
energy.

The large HOMO-LUMO gaps of Au$_{40}$ are related to
tetrahedral symmetry.
In perfectly spherical clusters
one finds each DFT Kohn-Sham orbital to be in an unique
angular momentum eigenstate relative to the
cluster's center of mass.
Due to the deformation of the nuclear background
away from spherical symmetry these states are not clean anymore.
A tetrahedral deformation as present here
can be described by an effective potential of the form
\cite{Hamamoto91zpd,Schnuck04prc}
\begin{equation}
  \label{eq:V}
  V({\bf r})=V(r) \left[ 1 + \alpha_{32} (T_{3,+2} -  T_{3,-2)})\right]
\end{equation}
where $r=|{\bf r}|$, the $T_{3,\pm 2}$ are spherical tensor operators and
$\alpha_{32}$ is a constant
describing the degree of the deformation. The exact form of
the $T_{3,\pm 2}$ is not important for our purpose; one only has to note that
these operators couple angular momentum eigenstates with angular
momentum projections that differ by $\pm 2$ exclusively.

\begin{figure}[ht]
  \epsfig{file=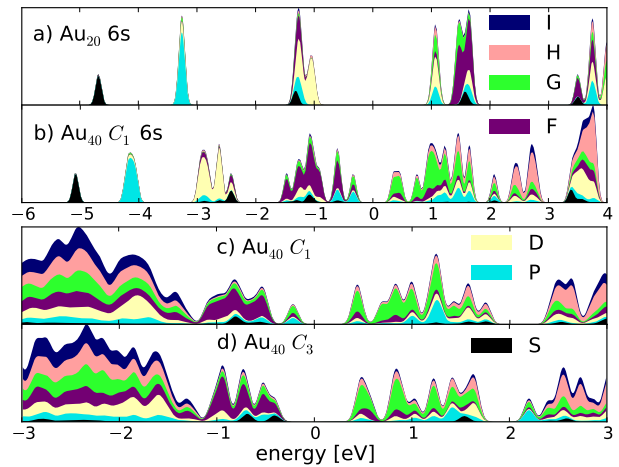, width=\linewidth}
  \caption{
    \label{fig:yl}
    (Color online) 
    The density of states relative to Fermi energy projected on
    angular momentum eigenfunctions relative to the clusters center
    of mass. In a, b) only the Au(6s) electrons
    are treated as valence electrons, whereas
    in c, d) also the Au(5d) electrons are allowed to rearrange.
  }
\end{figure}
With these considerations we analyze the electronic structure
of the tetrahedral gold clusters in Fig. \ref{fig:yl}
\footnote{The GPAW package \cite{Mortensen05prb, Enkovaara10jpc}
performing the PAW method on real space grids with 0.02 nm grid spacing
was used for this purpose}.
We project the DFT Kohn-Sham orbitals onto
spherical angular momentum eigenstates relative to the
cluster center of mass to extract their
delocalized nature \cite{Walter08pnas}.
The usual nomenclature characterizing the
states' angular momentum is similar to atomic physics $S, P, D, \dots$,
where capital letters  distinguish from atom
centered angular momenta.
The principle quantum number $n$ gives the energetic ordering
which is connected to the ($n-1$) number of radial nodes of the
corresponding eigenfunctions. In spherical clusters
one would expect the delocalized
states to fill 1S$^2$1P$^6$1D$^{10}$2S$^2$ orbitals for 20 electrons
and additionally the 1F$^{14}$2P$^6$ orbitals for 40 electrons.

To obtain a clean picture of delocalized states
we first fix all the electrons except the Au(6s) electrons
in a frozen core approximation.
The projected density of states (PDOS) for Au$_{20}$ obtained in this way
is presented in Fig.~\ref{fig:yl}a.
The states of lowest energy are the delocalized 1S
and three 1P states, similar as in the
spherical case.
This is understandable as the magnetic quantum number
of these states is 0, $\pm1$ and hence these could only couple to higher
angular momentum states far away in energy through the $T_{3,\pm 2}$
operators in eq. (\ref{eq:V}). Then
there follows a block of 4 states with mixed S/P/D/F symmetry
due to the $T_{3,\pm 2}$ coupling, immediately followed by two states of pure
D symmetry that form the Au$_{20}$ HOMO. After the substantial
gap of 2.06 eV, the Au$_{20}$ LUMO consists of
3 states with S/P/G symmetry and is
followed by a
block of 3+2 states of dominant F symmetry. After these there is
another large gap. This analysis clearly shows how sparse the
delocalized electronic states are distributed under
tetrahedral deformations and that the electronic system can make
gain from closing shells in the corresponding symmetry.

Analyzing the Au$_{40}$ states in the same way leads to a very comparable
picture as shown in Fig.~\ref{fig:yl}b.
The relative gaps between the blocks of
states gets smaller due to the larger size of the cluster, but the
symmetry of the states is similar. In this cluster the rather
large HOMO-LUMO gap is between the F/P symmetry dominated
occupied states and the G symmetry dominated unoccupied states.
Including Au(5$d$) states as valence electrons into the calculation does
not change the picture of the frontier orbitals around the Fermi
energy as shown in Fig.~\ref{fig:yl}c and d.
A comparison with the $C_3$ isomer that has a clearer tetrahedral
structure indicates that
in terms of symmetry
the HOMO of the $C_1$ isomer belongs rather to
the LUMO block of states with dominant P/G symmetry.
As a consequence the gap of the $C_3$ isomer is even
larger than that of
the $C_1$ isomer.
Finally we have analyzed the deformation of the $s$-valence 
electron density as it
was done for the near tetrahedral shape of Na$_{40}$  
\cite{Rytkonen98prl}. While we obtain for the largest
distortion parameter $S_3=0.04$ 
for Na$_{40}$ \cite{Huber09prb} in agreement
with ref. \onlinecite{Rytkonen98prl},
$S_3=0.26, 0.33$ for the
$C_1, C_3$ isomers of Au$_{40}$ respectively, accounting for the
much larger tetrahedral distortion present in the gold clusters.

In summary, we found Au$_{40}$ to be a magic cluster
with a quasi-tetrahedral symmetry. It has a twisted pyramid structure
discovered from DFT-based basin hopping for global minimum search and
built upon the putative global minima of Au$_{38}$.
This cluster is a
manifestation of the enhanced stability due to the tetrahedral
symmetry, predicted both in nuclear structure and by the jellium
model. Analysis of the delocalized electrons in Au$_{40}$ confirms the
shell-closing picture by the tetrahedral symmetry, similar to that of
Au$_{20}$.
The delocalized $6s$ electrons and the complex energy
landscape for clusters such as Au$_{40}$ cannot be accurately
described by empirical potentials, thereby making DFT-based
global-minimum search a necessity.

This work was supported by the Division of Chemical Sciences,
Geosciences, and Biosciences, Office of Basic Energy Sciences,
U.S. Department of Energy. This research used resources of the
National Energy Research Scientific Computing Center, which is
supported by the Office of Science of the U.S. Department of Energy
under Contract No. DE-AC02-05CH11231.
M.W. acknowledges computational
resources from RZ J\"ulich and the local bwGrid, and funding from
Deutsche Forschungsgemeinschaft. 
We thank M. Moseler for providing the structure of Na$_{40}$. 


%

\end{document}